# Estimation of the number of biophotons involved in the visual perception of a single-object image: Biophoton intensity can be considerably higher inside cells than outside

I. Bókkon[1*] V. Salari [2,3] J.A. Tuszynski[4] I. Antal[5]

[1]Doctoral School of Pharmaceutical and Pharmacological Sciences, Semmelweis University, Hungary
[2]Kerman Neuroscience Research Center (KNRC), Kerman, Iran
[3]Afzal Research Institute, Kerman, Iran
[4]Department of Physics, University of Alberta, Edmonton, T6G 2J1, Canada
[5]Department of Pharmaceutics, Semmelweis University, Budapest, Hungary
[*]Corresponding author: I. Bókkon. Contact: bokkoni@yahoo.com

**Abstract**

Recently, we have proposed a redox molecular hypothesis about the natural biophysical substrate of visual perception and imagery (Bókkon, 2009. BioSystems; Bókkon and D'Angiulli, 2009. Bioscience Hypotheses). Namely, the retina transforms external photon signals into electrical signals that are carried to the V1 (*striate cortex*). Then, V1 retinotopic electrical signals (*spike-related electrical signals along classical axonal-dendritic pathways*) can be converted into regulated ultraweak bioluminescent photons (*biophotons*) through redox processes within retinotopic visual neurons that make it possible to create intrinsic biophysical pictures during visual perception and imagery. However, the consensus opinion is to consider biophotons as by-products of cellular metabolism. This paper argues that biophotons are not by-products, other than originating from regulated cellular radical/redox processes. It also shows that the biophoton intensity can be considerably higher inside cells than outside. Our simple calculations, within a level of accuracy, suggest that the real biophoton intensity in retinotopic neurons may be sufficient for creating intrinsic biophysical picture representation of a single-object image during visual perception.

*Keywords:* Neurons, Visual perception, Free radicals, Biophotons; Biophysical picture representation, Mitochondria,

## 1. Introduction

The homeothermic state has been suggested to make the development of explicit memory possible and to allow the brain to operate on pictures during informational processing due to the regulated electrical and biophotonic mechanisms [1,2,3]. It was also suggested that the phosphene lights are the result of the intrinsic perception of induced or spontaneous increased biophoton emission in cells in different parts of the visual system [4]. Recently, it was pointed out [5] that not only retinal phosphenes but also the discrete dark noise of rods can be due to the natural lipid oxidation related (free radical) bioluminescent photons in the retina. A redox molecular hypothesis [6,7] has been formulated about the natural biophysical substrate of visual perception and imagery (see Fig.1).

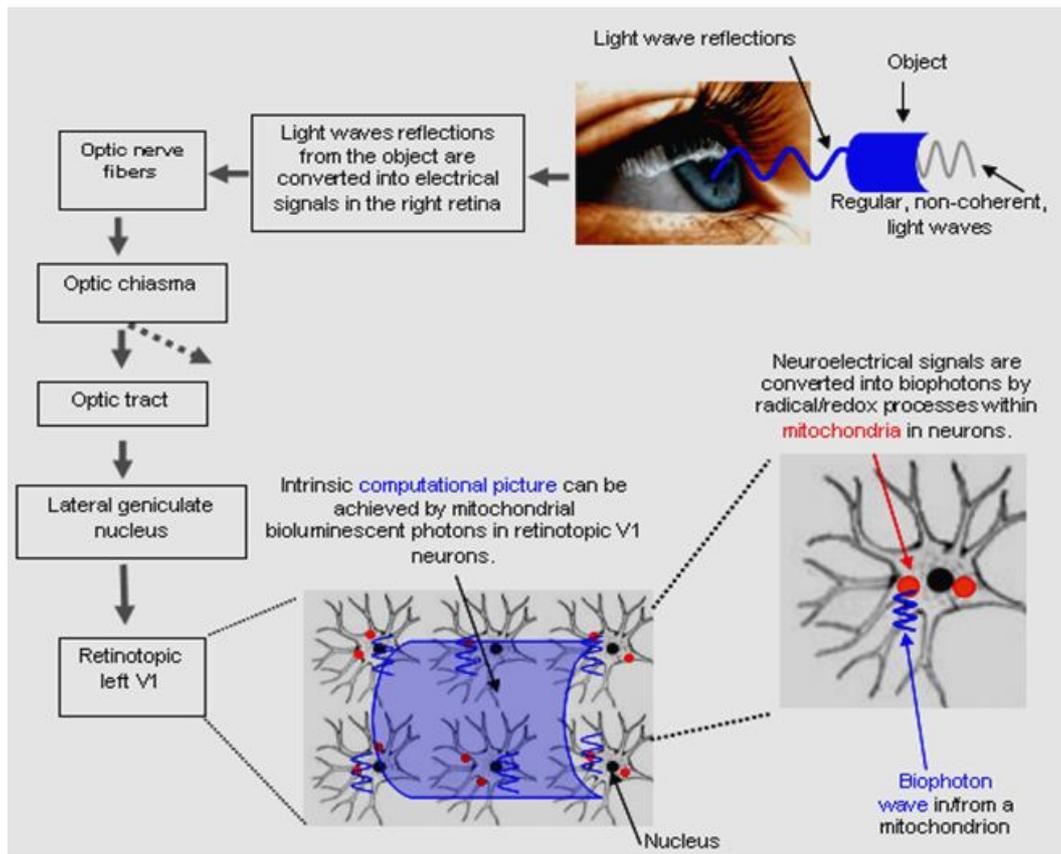

**Fig. 1.** Schematic representation of an intrinsic biophysical mechanism in visual perception (see [5] and [9] for details). Light waves from objects are converted into electrical signals in the retina. Retinotopical electrical signals are conveyed to V1 and converted into biophotons by mitochondrial redox processes in striate visual neurons. This gives an intrinsic computational picture of an object by bioluminescent biophotons in retinotopical V1. This model is limited to a static object.

Namely, the retina transforms photon signals from the external world into electrical (redox) signals that are conveyed to the V1 through the optic nerve. This V1 retinotopic electrical (*spike-related electrical signals along classical axonal-dendritic pathways*) information can be converted into spatio-temporal bioluminescent photon signals by mitochondrial and cellular redox processes that make it possible to generate intrinsic biophysical pictures in retinotopically organized mitochondrial cytochrome oxidase rich (CO-rich) visual areas during visual perception and imagery. In other words, the retinal photonic visual information can be approximately re-created by redox regulated biophotons of mitochondrial networks in visual neurons. However, if it can be demonstrated that perception of cortical phosphene light emission is due to neurocellular biophotons, intrinsic regulated biophotons of retinotopic visual areas can serve as a natural biophysical (*redox molecular*) substrate for visual perception and imagery.

It is noteworthy that Narici et al. [8] have supported the prediction [4, 8] about retinal phosphenes during space travel and stated that this is due to the ionizing radiation induced free radicals and chemiluminescent photons. In other words, ionizing radiation (*cosmic* *particles*) induced free radicals can create chemiluminescent photons from retinal lipid peroxidation. Subsequently, photons from retinal lipid peroxidation are absorbed by the photoreceptors, modify the rhodopsin molecules (*rhodopsin bleaching*) and initiate the phototransduction cascade resulting in the sensation of phosphene lights. Since retinal and cortical induced phosphenes are required to have a common molecular biophysical basis, the experiment of Narici et al. can be the first step toward proving our biophysical picture hypothesis.

It is worth noting that the term "ultraweak biophoton emission" can be misleading, since it suggests that biophotons are not important in cellular processes as mere by-products of cellular metabolism. In the following sections, we point out that biophotons originate from regulated redox/radical processes and the actual biophoton intensity can be fundamentally higher inside cells than outside. According to our rough estimation, at least $10^8$ - $10^9$ biophotons per second can be produced inside retinotopic visual neurons, which may be sufficient to create intrinsic biophysical picture representation during visual perception of a single-object image.

## 2. Controlled biophoton emission from free radical reactions

Free radical production and fundamentally unregulated (*stochastic*) process of aerobic oxidative metabolism have long been considered to be a health hazard. However, it is now clear that ROS (reactive oxygen species) and RNS (reactive nitrogen species) as well as their derivatives act as essential regulated signals in biological systems [9-12]. Cellular generation of ROS and RNS is vital for redox signaling. ROS-generating enzymes are compartmentalized [13] and strictly controlled at both the genetic and the activity levels [9].

ROS and RNS are produced mostly by the mitochondrial respiratory chain, NADPH oxidases, lipoxygenases, cyclooxygenases, cytochrome P450 oxidases, nitric oxide synthases, etc. [9,10]. ROS and RNS can regulate gene expression, apoptosis, cell growth, cell adhesion, chemotaxis, protein-protein interactions and enzymatic functions, $Ca^{2+}$ and redox homeostasis, and several other cellular processes [10, 13, 14-18].

There is experimental evidence that ROS and RNS are also essential for normal brain functions and synaptic processes. Free radicals and their derivatives act as signaling molecules in cerebral circulation and are essential in molecular signal processes such as synaptic plasticity, neurotransmitters release, memory formation, hippocampal long-term potentiation, etc. [19 -26].

During natural metabolic processes, in all types of living systems, lasting spontaneous photon emission has been detected without any external excitation [27-35]. Bioluminescent photon emission ranges from a few up to hundreds of photons per second per $cm^2$ within the spectral range of radiation from ultraviolet to near infrared. This ultraweak bioluminescent photon emission is referred to using various terms such as ultraweak photon, dark luminescence, low intensity chemiluminescence, ultraweak electromagnetic light, spontaneous autoluminescence, bioluminescence, biophotons, etc.

Biophotons originate from bioluminescent reactions of ROS and RNS and their derivatives, and also from simple cessation of electronically excited states. As examples we may list the mitochondrial respiration chain, lipid peroxidation, peroxisomal reactions, oxidation of catecholamines, oxidation of tyrosine and tryptophan residues in proteins etc. [36-39]. One of the major sources of biophotons is derived from mitochondrial oxidative metabolism (Fig.2) and lipid peroxidation.

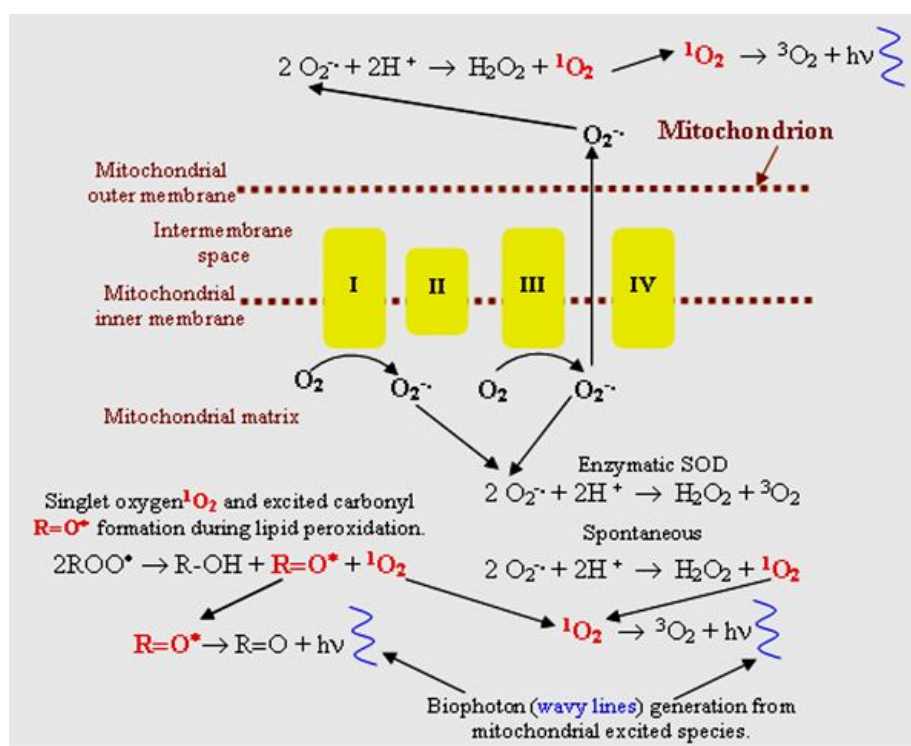

**Fig. 2.** Mitochondrial redox reactions are major sources of reactive and excited species. Biophotons originate from excited species. Major biophoton emission is due to the excited electrons of singlet oxygen $^1O_2$ and carbonyl species R=O*. When an excited carbonyl or singlet oxygen is released to the ground state, it can give out its energy as a photon (biophoton). Biophotons originate from regulated redox/radical processes, and the actual biophoton intensity can be radically higher inside cells than outside.

Since the generation of ROS and RNS is not a random process, but rather a precise mechanism used in cellular signaling pathways, the biophoton emission can also be a regulated process under both physiological and pathophysiological circumstances. In other words, regulated generation of ROS and RNS can lead to regulated biophoton generation in/from various cells during natural oxidative metabolism.

Biophotons can be absorbed by natural photosensitive chromophores of cells. For example, the electron transport chains on the inner membrane of mitochondria contain photosensitive chromospheres (*flavinic and pyridinic rings, porphyrin ring*) [39,40]. Fluorescent lipid chromophores can also act as photo-acceptors formed during regulated lipid peroxidation of membranes [41]. Photosensitive chromophore molecules of cells can transfer the absorbed biophoton energy to nearby biomolecules via resonance energy transfer, which can induce conformation changes and trigger complex signal processes in cells.

**3. Biophoton emission and neural activity**

There is a direct relationship between the intensity of biophotons and neural metabolic activity in rat hippocampal slices [27]. This biophoton emission can be facilitated by the membrane depolarization of neurons by a high concentration of $K^+$, and can be attenuated by tetrodotoxin or elimination of extracellular $Ca^{2+}$ [42]. In *in vivo* experiments, the spontaneous biophoton emission from a rat's brain is associated with cerebral energy metabolism, EEG activity, and oxidative stress [28, 43]. Biophoton intensity from the brain slices depends on temperature and oxygen concentration. Thus, we conclude that neuronal biophoton emission is in direct relationship with neural activity and neurobiochemical processes. Moreover, it was demonstrated that biophotons (ultraweak bioluminescent photons) can conduct along the neural fibers [44].

Since regulated electrical (*redox*) signals of neurons can be converted into regulated biophoton signals, external information representation can emerge not only as electrical signals but also as regulated biophoton (*ultraweak optical*) signals in the brain.

According to [6,7] and Figure1, the retina transforms photon signals from the external visual world into electrical signals that are conveyed to the V1. This retinotopic electrical information in V1 (*spike-related electrical signals along classical axonal-dendritic pathways*) can be converted into regulated spatio-temporal bioluminescent biophoton patterns by mitochondrial redox processes that make it possible to generate intrinsic biophysical pictures in mitochondrial CO-rich visual areas during visual imagery and visual perception. In other words, the retinal photonic visual information can be re-represented by redox regulated biophotons of mitochondrial networks in mitochondrial CO-rich visual areas.

Since externally measured biophoton intensity is extremely low, a counter-argument could easily be made. The question is how this very low biophoton intensity is capable of creating biophysical pictures in retinotopic mitochondrial CO-rich visual areas during visual perception and imagery. However, our biophysical hypothesis [6,7] is based on intrinsic regulated biophotons of visual neurons and not on the externally measured ultraweak biophoton emission of cells. According to our hypothesis, small clusters of visual neurons act as "visual pixels" appropriate to the topological distribution of photon stimuli on the retina.

In the next sections, we indicate that the real biophoton intensity within visual neurons may be sufficient for creating an intrinsic biophysical picture of a single-object image in retinotopic visual areas during visual perception.

## 4. Evaluation of the number of possible reactive oxygen species within cells

Specialized cells have very different oxygen consumption and they produce very different percentages of partially reduced oxygen molecules. For instance, neutrophils, may show a lot of activity while fibroblasts very little.

According to some calculations, roughly $10^{12}$ oxygen molecules are processed by each rat cell daily, and the leakage of partially reduced oxygen molecules is around $2\%$, yielding approximately $2 \times 10^{10}$ superoxide $O_2^-$ and $H_2O_2$ molecules per cell per day [45-47]. Under basal conditions, human cells can produce about one tenth the ROS of those in rats, or 2 billion ($2 \times 10^9$) superoxide and $H_2O_2$ molecules per cell per day [46,47].

Other calculations indicate that the average cell utilizes $10^{13}$ molecules of $O_2$ per cell per day which equates to $10^{11}$ free radical species produced per cell per day [48,49]. So, each brain cell is estimated to produce more than $10^{11}$ free radicals per day [49].

It is believed that under normal physiological conditions, only a small percentage of oxygen molecules causes ROS production. However, NADPH oxidases are ubiquitous enzymes that can perform direct oxygen reduction even under resting condition, and up to 20% of all consumed oxygen is directly reduced and goes to ROS production [50].

However, it is clear that there are substantial discrepancies regarding the calculated number of possible reactive oxygen species within cells as can be attested by papers stating estimated numbers of free radical species in cells without providing any variability or standard deviations [46-49]. Various cells exhibit very different oxygen consumption and produce very different percentages of partially reduced oxygen molecules. Moreover, the oxygen consumption and the number of partially reduced oxygen molecules in the same cell can change continuously within the cell cycle and in response to intrinsic and extrinsic factors. This goes to show that variability or standard deviation in the estimated number of free radical species in cells is not practical in this case.

A day has $8.64 \times 10^4$ seconds. The number of free radicals that can be produced in one brain cell during one day is $10^{11}$ [49] (Doraiswamy et al., Lancet Neurology, 2004), so dividing it by the number of seconds in a day yields $\frac{10^{11}}{8.64 \times 10^4} \Rightarrow 1,157,407$. This means that 1,157,407 free radicals may be processed per cell per second in one brain cell. Since the main source of biophotons originates from free radical reactions, this indicates that the actual number of biophotons - inside cells - should be significantly larger than that expected from biophoton measurements, which are usually carried out at a distance of several centimeters from the cells. It is probable that living cells retain their biophotons within the cellular environment for signal processing. Grass and Kasper's experiments [51] indicate that living systems are able to retain photons. For example, they reported that whole blood does not re-emit any biophotons after illumination. In contrast, the same measurement for serum gives intense biophoton re-emission for more than 10 minutes. According to Cliento [52], biochemical reactions take place in such a way, that a biophoton is borrowed from the surrounding electromagnetic bath, and then it excites the transition state complex, and finally returns to an equilibrium state with the surroundings.

**5. The number of possible biophotons within cells**

In Section 8, considerable discrepancies were discussed regarding the number of possible reactive oxygen species within cells. Since the number of biophoton generations within cells cannot be measured, it should not be surprising that it is almost impossible to estimate the real biophoton emissions in various living cells. So, the question is: what can be precisely measured in the cell? The answer is the actual biophoton intensity which can be considerably higher inside the cells than outside. It is very probable that externally measured ultraweak biophoton emission from various cells originates primarily from the natural oxidation processes of cellular membranes' surface areas.

According to Thar and Kühl [39], the real biophoton intensity within cells can be significantly higher than the one expected from the measurements of ultraweak bioluminescence (usually measured some centimeters away from the cells). Since photons are strongly scattered and absorbed in cellular systems, the corresponding intensity of biophotons within the organism can even be two orders of magnitude higher [53, 54].

Based on the data from the rat's brain biophoton experiments [28,43,55,56], we infer that $100 \frac{biophoton}{cm^2 \times sec}$ can be expected from the cortex surface. However, neurons absorb each other's emitted biophotons in a volume arrangement, which drastically decreases the number of measurable biophotons. Moreover, the real biophoton intensity inside cells can be two orders of magnitude higher than outside.

## 6. Estimates of the number of possible reactive oxygen species and biophotons during a single-object image

According to Pakkenberg and Gundersen [57], the neuronal density in V1 is $60000 \frac{neurons}{mm^3}$ (or $60 million \frac{neuron}{cm^3}$) in postmortem human brains. Since the V1 thickness is about 0.2 cm, and V1 surface area of one hemisphere is about 26 cm$^2$ in humans [58], this gives a 5.2 cm$^3$ ($26 cm^2 \times 0.2 cm = 5.2 cm^3$) volume of one hemisphere V1. As a result, we estimate that there are $5.2 cm^3 \times 60 million \frac{neurons}{cm^3} \cong 312 million\ visual\ neurons$ in V1 of one hemisphere.

According to Levy et al. [59], at least a million ($10^6$) neurons in object-related areas and at least 30 million neurons ($30 \times 10^6$ neurons) in the entire visual cortex are activated by a single-object image.

Let us use our calculated 1,157,407 free radicals that can be produced by each brain cell per second, which yields $10^6 neurons \times 1,157,407\ free\ radicals \approx 1.157 \times 10^{12} free\ radicals$. Namely, this represents $1.157 \times 10^{12}$ various free radicals produced by human visual neurons per second in V1 of one hemisphere during perception of a single-object image.

Since biophotons originate from free radical biochemical reactions, it also indicates that the externally measured biophoton emission - $\approx 100 \frac{biophoton}{cm^2 \times sec}$ - from the cortex surface is not able to provide evidence for the real biophoton emission inside neuronal cells.

Now, let us propose based on earlier estimates that at least 100 biophotons can be produced within each human visual neuron per second, and use $10^6$ visual neurons that are activated by a single-object image in object-related neurons which yields

$$\frac{1\,million\,visual\,neurons \times 100\,biophoton}{neuron \times \sec} \Rightarrow 10^8\,biophoton/\sec \text{ during a single-object image}$$

representation.

Since at least 30 million neurons in the entire visual cortex are activated by a single-object image, this results in

$$\frac{30\,million\,visual\,neurons \times 100\,biophoton}{neuron \times \sec} \Rightarrow 3 \times 10^9\,biophoton/\sec$$

during a single-object image representation per second.

Since the striate cortex (V1) and many extrastriate visual cortical areas are organized in a retinotopic manner [60,61], with regard to the topological distribution of photon stimuli on the retina, and considering small groups of visual neurons as "pixels", production of $10^8$ - $10^9$ biophoton within retinotopic visual neurons per second may be sufficient to generate an intrinsic biophysical picture representation during visual perception of a single-object image via neuro-computational processes.

## 7. Measuring internal biophotons with an ideal biophoton device

Kobayashi et al [28] have presented a biophoton device to measure the power intensity of biophoton emission from the rat's brain tissue. Their formulation for minimum detectable radiant flux $P_{min}$ ($W/cm^2$) is:

$$P_{min} = \frac{hc}{\lambda} \frac{\sigma_d}{\eta(\Omega/2\pi)\sqrt{TA}} \qquad (1)$$

where $\Omega$ is the solid angle for the light collection efficiency of the lens system. $\eta$ is the quantum efficiency of the photocathode at wavelength $\lambda$, $\sigma_d$ is the standard deviation of the dark count within the $T$ (observation time for integration, here $T=1$), $A$ is the active area of the detector, $c$ is the velocity of light and $h$ is Plank's constant. The relation (1) is obtained when minimum detectable optical power is defined as unity of the signal-to-noise ratio. The number of measured biophotons by device can be determined as:

$$N_B = \frac{hc}{\lambda} \frac{\sigma_d \sqrt{TA}}{\eta(\Omega/2\pi)k(\lambda)} \qquad (2)$$

where $N_B$ is denoted as the number of biophotons, and $k(\lambda)$ is a coefficient which can be a function of the wavelength $\lambda$. We have obtained eq. (2) from the relation $P_{min} = k(\lambda)\frac{N_B}{TA}$. Now, consider an ideal device to measure the real numbers of biophotons. First, we introduce the relation $N_B = N_I \exp(-\mu x)$, for the number of biophotons inside the tissue $N_I$ and outside the tissue $N_B$ where $\mu$ is the biophoton absorption coefficient of the cell and x is the thickness of the tissue cell membranes. Thus, substituting $N_I$ into eq. (2) from the relation above, we have,

$$N_I = \frac{hc}{\lambda} \frac{\sigma_d \sqrt{TA}}{\eta(\Omega/2\pi)k(\lambda)} \exp(\mu x) \qquad (3)$$

For an ideal biophoton device ($\sigma_d = 1, \eta = 1, \Omega/2\pi \approx 0.5$), for $T = 1s$ and for all produced biophotons inside the visual cortex for seeing a single object the relation becomes

$$N_I = 2\frac{hc}{\lambda}\frac{\exp(\mu x)}{k(\lambda)} \qquad (4)$$

where $k(\lambda)$ is a Gaussian function which can be written in the form of $k(\lambda) = \beta \exp(-\frac{\lambda^2}{2\xi^2})$ where $\beta$ and $\gamma$ can be obtained from the equation:

$$\int_{-\infty}^{\infty} \beta e^{-\frac{\lambda^2}{2\xi^2}} d\lambda = \beta\xi\sqrt{2\pi}. \qquad (5)$$

The parameter $\xi$ is related to the full width at half maximum of the peak, $2\sqrt{2\ln 2}\xi \approx 2.3\xi$. For example, for a $\lambda = 630 nm$ in peak and width of *100 nm*, $\xi \approx 50$ [28], so $\beta$ can be determined from this result that when for $\lambda = 630 nm$, $P_m = 8 \times 10^{-17} (W/cm^2)$, it equals to $2.5 \times 10^2 \frac{photon}{s.cm^2}$, thus $k(\lambda) = 3.2 \times 10^{-19}$, hence $\beta = 9.5 \times 10^{15}$, consequently $k(\lambda) = 9.5 \times 10^{15} \exp(-\frac{\lambda^2}{2\xi^2})$.

To investigate the number of biophotons produced inside the visual cortex, we use eq. (4) and find

$$N_I \approx \frac{4 \times 10^{-9}}{\lambda} \exp(\mu x + \frac{\lambda^2}{2\xi^2}) \qquad (6)$$

where $\mu x$ in eq. (6) can be obtained from the relation $N_B = N_I \exp(-\mu x)$, putting in eq. (6) with the assumption that $\tau = \frac{\lambda}{\xi}$ and $\mu x = \ln(\frac{N_I}{N_B})$,

$$N_I \approx \frac{4}{\lambda} \exp(\mu x + \frac{\tau^2}{2}) \qquad (7)$$

where $\lambda$ is measured in nanometers. Figure 3 represents the number of internal biophotons in terms of $\mu x$ and $\tau$. The perpendicular axis on the horizontal plane is related to the number of biophotons produced in the visual cortex required for seeing a single object.

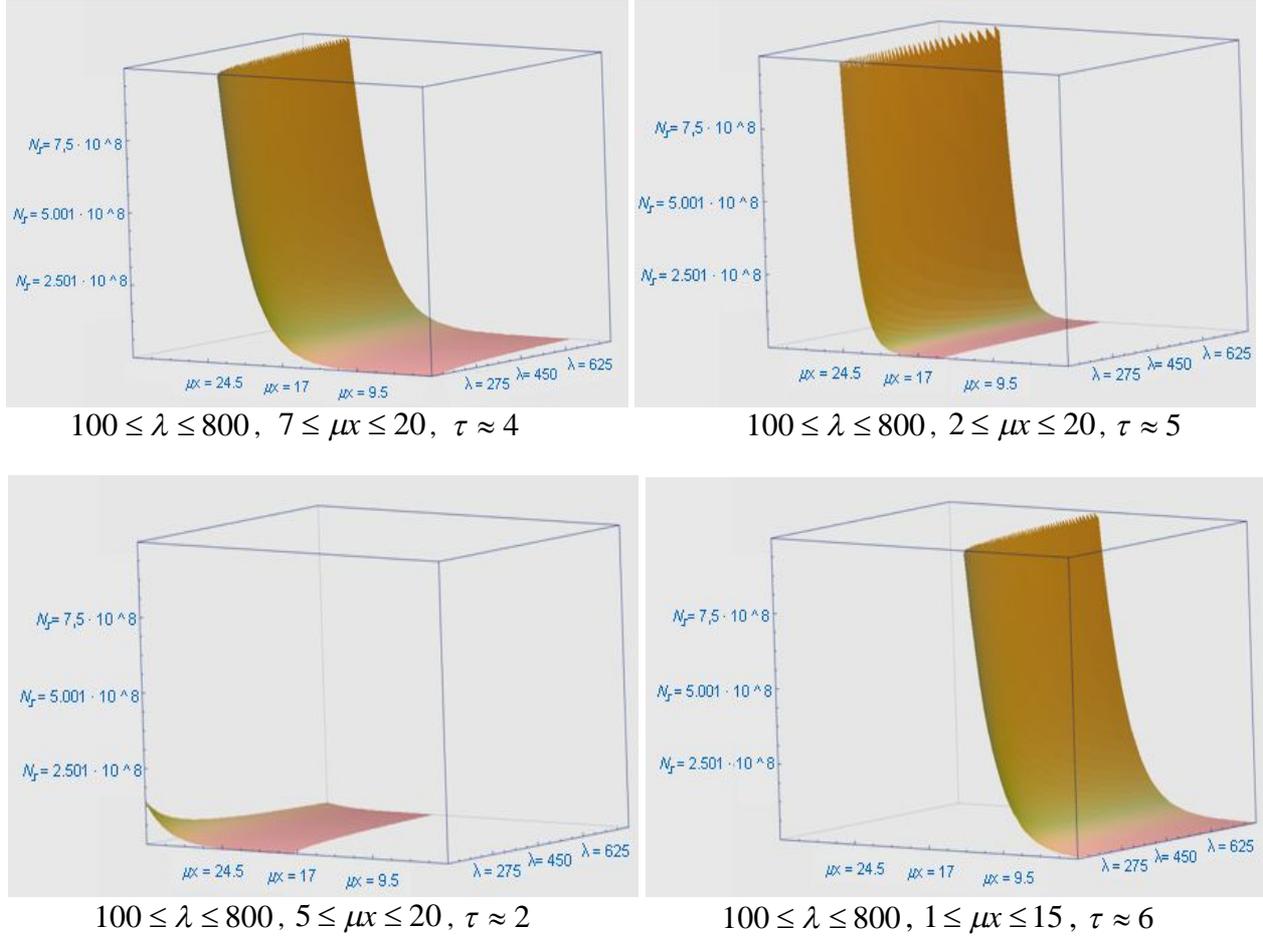

**Fig. 3**. Representation of internal biophoton production for different values of $\mu x$ and $\tau$.

It is seen in Fig.3 that $\mu x$ or $\ln(\frac{N_I}{N_B})$ must be on the order of 10 to have internal biophotons. These diagrams show that the number of internal biophotons is zero for other measures and else their amount is very high on the order of $10^8 - 10^9$ biophotons. This result is in accordance with our earlier conclusions. The power of internal biophotons can be obtained by relation $P_{real} = P_{measured} e^{\mu x}$ in which $P_{measured}$ is the power measured by real biophoton device with low efficiency and $P_{real}$ is the real power of internal biophotons. According to our diagrams, $2 \leq \mu x \leq 20$, the amount of $P_{real}$ is in the range

$$e^2 \times 10^{-17} (W/cm^2) \leq P_{real} \leq e^{20} \times 10^{-17} (W/cm^2)$$
$$\rightarrow \quad 10^{-16}(W/cm^2) \leq P_{real} \leq 10^{-7}(W/cm^2)$$

## 8. Mitochondria and centrioles

In a series of studies spanning a period of some 25 years G Albrecht-Buehler (AB) demonstrated that living cells possess a spatial orientation mechanism located in the centriole [62-64]. This is based on an intricate arrangement of microtubule filaments in two sets of nine triplets each of which is perpendicular to the other. This arrangement provides the cell with a primitive "eye" that allows it to locate the position of other cells within a two to three degree accuracy in the azimuthal plane and with respect to the axis perpendicular to it [64]. He further showed that electromagnetic signals are the triggers for the cells' repositioning. It is still largely a mystery how the reception of electromagnetic radiation is accomplished by the centriole. Another mystery related to these observations is the original electromagnetic radiation emitted by a living cell [65]. Using pulsating infra-red signals scattered off plastic beads AB mimicked the effects of the presence of another living cell in the neighbourhood. The question that still remains unanswered and which we address here is the source of infra-red radiation speculated by AB to originate in the mitochondria and later on demonstrated to be correct using quantum mechanical arguments [65]. Mitochondria are not only the main source of bioluminescent biophotons generated by regulated reactive oxygen species, but they also function as photoreceptors with their natural photosensitive chromophores [66,67]. Ultraweak biophotons can be guided along the mitochondrial network [39]. Thus, the mitochondrial network can work as a bioluminescent biophoton communication network in cells and neurons.

Vertebrate cells are in general filamentous mitochondria associated with the microtubules of the cytoskeleton, forming together a continuous network (mitochondrial reticulum) [68,69]. Mitochondrial movements in both axons and dendrites are mainly microtubule-driven, but in each domain there may also be some movement along actin filaments. The fast movement of mitochondria is microtubule-based but the slower movement is actin-based [70].

Both mitochondria and microtubules can form dynamic networks in neurons. Moreover, the refractive index of both mitochondria and microtubules is higher than the surrounding cytoplasm, whose consequence is that mitochondria and microtubules can act as optical waveguides, i.e. electromagnetic radiation (*biophotons*) can propagate within their network [39]. Regulated biophotons (*from mitochondrial radicals and excited molecules*) can induce polymerization of microtubules. Then, according to the quality of absorbed biophotons from mitochondria, microtubules can transport mitochondria in accordance with information processes in cells and neurons. There can be a mutual cross-talk/regulation between mitochondria and microtubules by redox and free radical processes. We speculate on the basis of our arguments presented in this paper that in these processes the "choirmaster" is the mitochondrial network.

**9. Summary**

In this paper we have presented an argument that biophotons are not by-products, but originate from regulated redox/radical processes. Moreover, we have stressed that the actual biophoton intensity can be drastically higher inside cells compared to their surrounding environment. According to our approximate calculations, the real biophoton intensity within retinotopic visual neurons may be sufficient to produce intrinsic biophysical picture representation of a single-object image during visual perception. In other words, the retinal photonic visual information can be re-represented by redox regulated biophotons in retinotopic visual neurons during visual perception. There also can be a mutual cross-talk/regulation between mitochondria and microtubules by redox and free radical processes. It is highly feasible for living cells and neurons to retain their biophotons within the cellular environment for use in biophysical signal processes. The major challenge, however, is to develop experimental techniques which would make it possible to accurately measure the levels of biophoton intensity inside living cells, especially neurons.

**Declaration of interest**

The authors report no conflicts of interest. The authors alone are responsible for the content.

**Acknowledgements**

Bókkon I. gratefully acknowledges support of this work by the BioLabor (Hungary), www.biolabor.org; Bókkon's URL: http://bokkon-brain-imagery.5mp.eu
Salari V. thanks Dr. Eskandary H., Dr. Sheibani V. and Dr. Rahnama M. for their scientific helps.

**References**
[1]  I. Bókkon, Creative Information, J. Biological Systems 1 (2003)1-17.
[2]  I. Bókkon, Dream pictures, neuroholography and the laws of physics, J. Sleep Res. 15 (Suppl. 1) (2006)187.
[3]  I. Bókkon, Dreams and Neuroholography: An Interdisciplinary Interpretation of Development of Homeotherm State in Evolution, Sleep Hypnosis 7 (2005) 61-76.
[4]  I. Bókkon, Phosphene phenomenon: A new concept, BioSystems 92 (2008)168-174.
[5]  I. Bókkon, RLP. Vimal, Retinal phosphenes and discrete dark noises in rods: a new biophysical framework. J. Photochem. Photobiol. B: Biology. 96 (2009) 255-259.
[6]  I. Bókkon, A. D'Angiulli, Emergence and transmission of visual awareness through optical coding in the brain: A redox molecular hypothesis on visual mental imagery, Bioscience Hypotheses 2 (2009) 226-232.
[7]  I. Bókkon, Visual perception and imagery: A new molecular hypothesis, BioSystems 96 (2009) 178-184.
[8]  L. Narici, A. De Martino, V. Brunetti, A. Rinaldi, W.G. Sannita, M. Paci, Radicals excess in the retina: A model for light flashes in space, Radiation Measurements 44 (2009) 203-205.

[9] K. Bedard, K.H. Krause, The NOX family of ROS-generating NADPH oxidases: physiology and pathophysiology, Physiol. Rev. 87 (2007) 245-313.

[10] W. Dröge, Free Radicals in the Physiological Control of Cell Function, Physiol. Rev. 82 (2002) 47-95.

[11] H.J. Forman, J.M. Fukuto, T. Miller, H. Zhang, A. Rinna, S. Levy, The chemistry of cell signaling by reactive oxygen and nitrogen species and 4-hydroxynonenal, Arch. Biochem. Biophys. 477 (2008) 183-195.

[12] M. Valko, D. Leibfritz, J. Moncol, M.T. Cronin, M. Mazur, J. Telser, Free radicals and antioxidants in normal physiological functions and human disease, Int. J. Biochem. Cell. Biol. 39 (2007) 44-84.

[13] L.S. Terada, Specificity in reactive oxidant signaling: think globally, act locally, J. Cell. Biol. 174 (2006) 615-623.

[14] P. Chiarugi, Reactive oxygen species as mediators of cell adhesion, Ital. J. Biochem. 52 (2003) 28-32.

[15] P. Chiarugi, Src redox regulation: there is more than meets the eye, Mol. Cells 26 (2008) 329-337.

[16] R.F. Feissner, J. Skalska, W.E. Gaum, S.S. Sheu, Crosstalk signaling between mitochondrial Ca2+ and ROS, Front. Biosci. 14 (2009) 1197-1218.

[17] A.V. Gordeeva, R.A. Zvyagilskaya, Y.A. Labas, Cross-talk between reactive oxygen species and calcium in living cells, Biochemistry (Mosc) 68 (2003) 1077-1080.

[18] J.T. Hancock, R. Desikan, S.J. Neill, Role of reactive oxygen species in cell signalling pathways, Biochem. Soc. Trans. 29 (2001) 345-350.

[19] C. Hidalgo, M.A. Carrasco, P. Muñoz, M.T. Núñez, A role for reactive oxygen/nitrogen species and iron on neuronal synaptic plasticity, Antioxid. Redox Signal. 9 (2000) 245-255.

[20] A. Kamsler, M. Segal, Control of neuronal plasticity by reactive oxygen species, Antioxid. Redox Signal. 9 (2007)165-167.

[21] K.T. Kishida, E. Klann, Sources and targets of reactive oxygen species in synaptic plasticity and memory, Antioxid. Redox Signal. 9 (2007) 233-244.

[22] L.T. Knapp, E. Klann, Role of reactive oxygen species in hippocampal long-term potentiation: contributory or inhibitory? J. Neurosci. Res. 70 (2002)1-7.

[23] M.V. Tejada-Simon, F. Serrano, L.E. Villasana, B.I. Kanterewicz, G.Y. Wu, M.T. Quinn, E. Klann, Synaptic localization of a functional NADPH oxidase in the mouse hippocampus, Mol. Cell. Neurosci. 29 (2005) 97-106.

[24] E. Thiels, E. Klann, Hippocampal memory and plasticity in superoxide dismutase mutant mice, Physiol. Behav. 77 (2002) 601-605.

[25] E. Thiels, N.N. Urban, G.R. Gonzalez-Burgos, B.I. Kanterewicz, G. Barrionuevo, C.T. Chu, T.D. Oury, E. Klann, Impairment of long-term potentiation and associative memory in mice that overexpress extracellular superoxide dismutase, J. Neurosci. 20 (2000) 7631-7639.

[26] A. Volterra, D. Trotti, C. Tromba, S. Floridi, G. Racagni, Glutamate uptake inhibition by oxygen free radicals in rat cortical astrocytes, J. Neurosci. 14 (1994) 2924-2932.

[27] Y. Isojima, T. Isoshima, K. Nagai, K. Kikuchi, H. Nakagawa, Ultraweak biochemiluminescence detected from rat hippocampal slices, NeuroReport 6 (1995) 658-660.

[28] M, Kobayashi, M, Takeda, K, Ito, H, Kato, H. Inaba, Two-dimensional photon counting imaging and spatiotemporal characterization of ultraweak photon emission from a rat's brain in vivo, J. Neurosci. Methods 93 (1999) 163-168.

[29] F.A. Popp, W. Nagl, K.H. Li, W. Scholz, O. Weingartner, R. Wolf, Biophoton emission. New evidence for coherence and DNA as source, Cell. Biophys. 6 (1984) 33-52.

[30] T.I. Quickenden, S.S. Que Hee, Weak luminescence from the yeast Sachharomyces-Cervisiae, Biochem. Biophys. Res. Commun. 60 (1964) 764-770.

[31] R.Q. Scott, P. Roschger, B. Devaraj, H. Inaba, Monitoring a mammalian nuclear membrane phase transition by intrinsic ultraweak light emission, FEBS Lett. 285 (1991) 97-98.

[32] M. Takeda, Y. Tanno, M. Kobayashi, M. Usa, N. Ohuchi, S. Satomi, H. Inaba, A novel method of assessing carcinoma cell proliferation by biophoton emission, Cancer Lett. 27 (1998) 155-160.

[33] R.N. Tilbury, T.I. Cluickenden, Spectral and time dependence studies of the ultraweak bioluminescence emitted by the bacterium Escherichia coli, Photobiochem. Photobiophys. 47 (1988) 145-150.

[34] R. Van Wijk, D.H. Schamhart, Regulatory aspects of low intensity photon emission, Experientia 44 (1988) 586-593.

[35] Y.Z. Yoon, J. Kim, B.C. Lee, Y.U. Kim, S.K. Lee, K.S. Soh, Changes in ultraweak photon emission and heart rate variability of epinephrine-injected rats, Gen. Physiol. Biophys. 24 (2005)147-159.

[36] I. Kruk, K. Lichszteld, T. Michalska, J. Wronska, M. Bounias, The formation of singlet oxygen during oxidation of catechol amines as detected by infrared chemiluminescence and spectrophotometric method, Z Naturforsch. [C] 44 (1989) 895-900.

[37] M. Nakano, Low-level chemiluminescence during lipid peroxidations and enzymatic reactions, J. Biolumin. Chemilum. 4 (2005) 231-240.

[38] B.P. Watts, M. Barnard, J.F. Turrens, Peroxynitrite-Dependent Chemiluminescence of Amino Acids, Proteins, and Intact Cells, Arch. Biochem. Biophys. 317 (1995) 324-330.

[39] R. Thar, M. Kühl, Propagation of electromagnetic radiation in mitochondria? J. Theor. Biol. 230 (2004) 261-270.

[40] T. Karu, Primary and secondary mechanisms of action of visible to near-IR radiation on cells, J. Photochem. Photobiol. B. 49 (1999) 1-17.

[41] V.M. Mazhul', D.G. Shcherbin, Phosphorescent analysis of lipid peroxidation products in liposomes, Biofizika 44 (1999) 676-681.

[42] Y. Kataoka, Y. Cui, A. Yamagata, M. Niigaki, T. Hirohata, N. Oishi, Y. Watanabe, Activity-Dependent Neural Tissue Oxidation Emits Intrinsic Ultraweak Photons, Biochem. Biophys. Res. Commun. 285 (2001) 1007-1011.

[43] M. Kobayashi, M. Takeda, T. Sato, Y. Yamazaki, K. Kaneko, K. Ito, H. Kato, H. Inaba, In vivo imaging of spontaneous ultraweak photon emission from a rat's brain correlated with cerebral energy metabolism and oxidative stress, Neurosci. Res. 34 (1999)103-113.

[44] Y. Sun, Ch. Wang, J. Dai, Biophotons as neural communication signals demonstrated by in situ biophoton autography, Photochem. Photobiol. 9 (2010) 315-322.

[45] B. Chance, H. Sies, A.B. Boveris, Hydroperoxide metabolism in mammalian organs, Physiol. Rev. 59 (1979) 527-605.

[46] J.R. Hoidal, Reactive Oxygen Species and Cell Signaling, Am. J. Respir. Cell. Mol. Biol. 25 (2001) 661-663.

[47] B.N. Ames, M.K. Shigenaga, T.M. Hagen, Oxidants, antioxidants, and the degenerative diseases of aging. Proc. Natl. Acad. Sci. USA 90 (1993) 7915-7922.

[48] G. Perry, L.M. Sayre, C.S. Atwood, R.J. Castellani, A.D. Cash, C.A. Rottkamp, M.A. Smith, The role of iron and copper in the aetiology of neurodegenerative disorders: therapeutic implications, CNS Drugs 16 (2002) 339-352.

[49] P.M. Doraiswamy, A.E. Finefrock, Metals in our minds: therapeutic implications for neurodegenerative disorders, Lancet Neurol. 3 (2004) 431-434.

[50] H.P. Souza, X. Liu, A. Samouilov, P. Kuppusamy, F.R. Laurindo, J.L. Zweier, Quantitation of superoxide generation and substrate utilization by vascular NAD(P)H oxidase, Am. J. Physiol. Heart Circ. Physiol. 282 (2002) H466-H474.

[51] F. Grass, S. Kasper, Humoral phototransduction: light transportation in the blood, and possible biological effects, Med. Hypotheses 71 (2008) 314-317.

[52] G. Cliento, Photobiochemistry without light, Experientia 44 (1988) 572-576.

[53] B.W. Chwirot, Ultraweak luminescence studies of microsporogenesis in Larch, in Popp FA, Li KH, Gu Q (eds.), Recent advances in biophoton research and its applications, Singapore, World Scientific Publishing Company, (1992) pp. 259-285.

[54] J. Slawinski, Luminescence research and its relation to ultraweak cell radiation, Experientia 44 (1988) 559-571.

[55] A.M. Adamo, S.F. Llesuy, J.M. Pasquini, A. Boveris, Brain chemiluminescence and oxidative stress in hyperthyroid rats, Biochem. J. 263 (1989) 273-277.

[56] S. Imaizumi, T. Kayama, J. Suzuki, Chemiluminescence in hypoxic brain--the first report, Correlation between energy metabolism and free radical reaction, Stroke 15 (1984) 1061-1065.

[57] B. Pakkenberg, H.J. Gundersen, Neocortical neuron number in humans: effect of sex and age, J. Comp. Neurol. 384 (1987) 312-320.

[58] D.L. Adams, L.C. Sincich, J.C. Horton, Complete pattern of ocular dominance columns in human primary visual cortex, J. Neurosci. 27 (2007) 10391-10403.

[59] I. Levy, U. Hasson, R. Malach, One picture is worth at least a million neurons, Curr. Biol. 14 (2004) 996-1001.

[60] A. Martínez, L. Anllo-Vento, M.I. Sereno, L.R. Frank, R.B. Buxton, D.J. Dubowitz, E.C. Wong, H. Hinrichs, H.J. Heinze, S.A. Hillyard, Involvement of striate and extrastriate visual cortical areas in spatial attention, Nat. Neurosci. 2 (1999) 364-369.

[61] T. Kaido, T. Hoshida, T. Taoka, T Sakaki, Retinotopy with coordinates of lateral occipital cortex in humans, J. Neurosurg. 101 (2004)114-118.

[62] G. Albrecht-Buehler, Cellular infrared detector appears to be contained in the centrosome., Cell Motil. Cytoskeleton 27 (1994) 262-271.

[63] G. Albrecht-Buehler, Changes of cell behavior by near-infrared signals. Cell Motil. Cytoskeleton, 32 (1995) 299-304.

[64] G. Albrecht-Buehler, Autofluorescence of live purple bacteria in the near infrared. Exp. Cell Res. 236 (1997) 43-50.

[65] J.A. Tuszynski, J.M. Dixon, Quantitative Analysis of the Frequency Spectrum of the Radiation Emitted by Cytochrome Oxidase Enzymes, Physical Review E, 64 (2001) 051915.

[66] M. Kato, K. Shinzawa, S. Yoshikawa, Cytochrome oxidase is a possible photoreceptor in mitochondria, J. Photochem. Photobiol. B: Biology 2 (1981) 263-269.

[67] T. Karu, Primary and secondary mechanisms of action of visible to near-IR radiation on cells, J. Photochem. Photobiol. B: Biology 49 (1999) 1-17.

[68] V.P. Skulachev, Mitochondrial filaments and clusters as intracellular power-transmitting cables, Trends Biochem. Scie. 26 (2001) 23-29.

[69] A.V. Kuznetsov, M. Hermann, V. Saks, P. Hengster, R. Margreiter, The cell-type specificity of mitochondrial dynamics, Int. J. Biochem. Cell Biol. 41 (2009) 1928-1239.

[70] L.A. Ligon, O. Steward, Movement of mitochondria in the axons and dendrites of cultured hippocampal neurons, J. Comp. Neurol. 427 (2000) 340-350.